\documentclass[preprint,12pt]{article}

\usepackage{arxiv}
\usepackage{titling}
\usepackage{setspace}
\usepackage{graphicx,color,psfrag}
\usepackage{fancyheadings}
\usepackage{url}

\usepackage[backend=biber,style=apa]{biblatex}
\bibliography{Homelessness Patterns}
\bibliography{Homelessness and COVID}
\bibliography{Network Visualization}
\bibliography{Me}

\title{A Graph Analysis of the Impact of COVID-19 on Emergency Housing Shelter Access Patterns}

\author{Geoffrey Messier  \\
  University of Calgary  \\
  2500 University Dr.~NW, Calgary, AB, Canada, T2N 1N4\\
  gmessier@ucalgary.ca
}

\date{}

\newcommand{\bE}{\begin{enumerate}}
\newcommand{\eE}{\end{enumerate}}
\newcommand{\bI}{\begin{itemize}}
\newcommand{\eI}{\end{itemize}}

\AtEveryBibitem{\clearfield{isbn}}
\AtEveryBibitem{\clearfield{issn}}

\begin{document}

\maketitle

\begin{abstract}
This paper investigates how COVID-19 disrupted emergency housing shelter access patterns in Calgary, Canada and what aspects of these changes persist to the present day.  This analysis will utilize aggregated shelter access data for over 40,000 individuals from seven major urban shelters dating from 2018 to the present.  A graph theoretic approach will be used to examine the journeys of individuals between shelters before, during and after the COVID-19 lockdown period.  This approach treats shelters as nodes in a graph and a person's transition between shelter as an arrow or edge between nodes.  This perspective is used to create both timeline and network diagrams that visualize shelter use and the flow of people between shelters.  Statistical results are also presented that illustrate the differences between the cohorts of people who only used shelter pre/post-lockdown, people who stayed in shelter during lockdown and people who used shelter for the first time during lockdown.  The results demonstrate not only how a complex system of care responded to the pandemic but also the characteristics of the people most likely to continue to rely on that system during an emergency.
\end{abstract}

\keywords{emergency shelter, housing first, vulnerability indicators, chronic homelessness, COVID-19 pandemic}

% --------------------------------------------------------------
\section{Introduction}
\label{sec:Intro}

Understanding how COVID-19 changed the movement of people through systems of care is important to gain a retrospective appreciation of the pandemic and to help plan for future similar events.  Now that many of the pandemic related restrictions have been lifted, it is equally important to understand which trends established during COVID-19 endure.  This later perspective is most urgent since it informs whether a system of care needs to adapt to more effectively address the long term effects of the pandemic.

During the initial onset of the COVID-19 pandemic, housing and homelessness systems of care (HHSCs) across the world transformed in an effort to adopt measures to limit the spread of the virus \parencite{oudshoorn2021, levesque2022, parsell2023, corey2022}.  This transformation was particularly urgent in the congregate living settings of emergency housing shelters.  The pandemic precipitated the opening of temporary additional spaces, an increased emphasis on housing placements to facilitate exits from shelter and a range of in-shelter measures designed to reduce disease spread.

The contribution of this paper is to illuminate how the pandemic and its associated impacts on shelter operations affected the flow of people through an HHSC in a major North American city.  This analysis will focus specifically on the emergency shelter component of that HHSC.  This is important since shelters typically have the lowest barrier to entry within an HHSC and often serve the people with the most complex physical and mental health challenges.  Shelters are also congregate living environments which make them particularly susceptible to viral spread \parencite{levesque2022}.

The data used in this paper captures the flow of 43,263 people between the top seven busiest emergency shelters in Calgary, Canada from March 1, 2018 to May 1, 2023.  Critically, this data provides not just shelter access information during the height of the pandemic but also for the two years leading up to the start of the pandemic and for approximately two years after most COVID-19 restrictions were eased.  This will demonstrate not only how the Calgary HHSC changed in response to the pandemic but also which of these changes continue to persist.

There have been only a limited number studies on how people flow between shelters due to the often fragmented nature of HHSC shelter data \parencite{culhane1994, jadidzadeh2020}.  Beyond these publications, shelter access patterns have mostly been analyzed by applying clustering techniques to stay and episode data for shelter use \parencite{kuhn1998, kneebone2015, aubry2013} which does not shed light on movement between shelters.  The work in \parencite{jadidzadeh2020} is particularly relevant since it was also done for the Calgary HHSC in response to the COVID-19 pandemic.  However, while COVID-19 motivated \parencite{jadidzadeh2020}, the study was conducted entirely using pre-pandemic data.  It is also limited to an investigation of how often a person chooses to move to a different shelter but does not provide information on which specific shelter has been chosen.  This is an important omission since understanding which paths between shelters see the most traffic is a critical part of efficiently allocating resources to support shelter users during an emergency or otherwise.

This omission can be addressed by creating a picture that shows not only when someone leaves a particular shelter but also where they have chosen to go.  This is not straightforward since the journey of each person through an HHSC is unique and often involves interacting with many services in different orders and for different lengths of time.  However, this paper will demonstrate that this complexity can be managed by representing the HHSC as a directed graph.   Graph theory and visualization is a vast field \parencite{bollobas1998} that has found utility in better understanding many diverse phenomena including the transit system, the Internet, social networks \parencite{derrible2011, majeed2020}. 

In the following, Section~\ref{sec:methods} will describe the data set used to study HHSC shelter use before, during and after the height of the COVID-19 lockdown restrictions.  It also describes how this data is pre-processed to create a directed graph that represents the flow of people through the HHSC shelter system.  This graph perspective is used to create timeline diagrams, cohort statistics and network diagram visualizations in Section~\ref{sec:results}.  Section~\ref{sec:discussion} will discuss the implications of these results and concluding remarks are made in Section~\ref{sec:conclusion}.

% --------------------------------------------------------------
\section{Methods}
\label{sec:methods}

% - - - - - - - - - - - - - - - - - - - - - - - - - - - - - - - - 
\subsection{The Data Set}
\label{ssec:data}

This study is conducted using daily emergency housing shelter access data collected and aggregated by the Calgary Homeless Foundation (CHF) between March 1, 2018 and May 1, 2023.  Records from multiple shelters for the same individual are linked and anonymized by the CHF before being released to the researchers.  Each data record consists of an anonymized individual identifier, shelter access date, shelter name and duration of the shelter stay.  The protocol governing the anonymization, secure storage and analysis of this secondary data set was approved by the University of Calgary Conjoint Ethics Review Board (REB-19-0095).  The data set contains 43,263 people with records of accessing an HHSC consisting of four adult shelters, two family shelters and one seniors' shelter.

To properly illustrate the effect of the pandemic, it is necessary to divide the data timeline into three periods.  The term ``lockdown'' will be used to refer to the period where shelter activity noticeably reduced due to COVID-19 restrictions.  The {\em pre-lockdown} period stretches from the start of the data on March 1, 2018 to March 17, 2020,  the {\em lockdown} period is from March 18, 2020 to July 1, 2021 and the {\em post-lockdown} period is from July 2, 2021 to May 1, 2023.  March 17, 2020 was the declaration of the first COVID-19 related state of emergency in Alberta \parencite{canadianinstituteforhealthinformation2022} and corresponds approximately to the introduction of widespread COVID-19 restrictions in Canada around the world.  July 1, 2021 corresponds to reaching Stage 3 of Alberta's ``Open for Summer'' plan \parencite{canadianinstituteforhealthinformation2022}.  

Note that it is not the intent of this analysis to declare COVID-19 ``over'' on July 1, 2021.  Clearly, both the spread of COVID-19, some pandemic related restrictions and the long term medical and societal effects of the pandemic continued to persist after this date.  July 1, 2021 is selected since it approximately marks a noticeable increase in shelter use activity, as will be illustrated in Section~\ref{sec:results}.

To more clearly understand the impact of COVID-19 on emergency shelter users, it is useful to divide the emergency shelter population into cohorts based on how their first and last days of shelter use co-incide with the COVID-19 lockdown periods.  These cohorts, the number of people in each cohort and the cohort inclusion criteria are shown in Table~\ref{tb.cohorts}.  A 30 day exclusion window relative to the start and end dates in the data set were used to calculate the dates in Table~\ref{tb.cohorts} in order to reduce the number of people included in the cohorts who have records of shelter access that are left or right censored by the start or end of the data set.

\begin{table}[htbp]
\centering
\begin{tabular}{c|c|p{3in}}
Cohort Name & Size & Criteria\\ \hline
Before Lockdown (Before) &  13654/43263 (31.6\%) & Started after 31/03/2018. Ended before 17/03/2020 \\
Stayed after Lockdown (Stayed) &  7986/43263 (18.5\%) & Started between 31/03/2018 and 17/03/2020. Ended between 31/03/2020 and 01/04/2023 \\
Started during Lockdown (During) &  3874/43263 (9.0\%) & Started between 17/03/2020 and 01/06/2021. Ended before 01/04/2023 \\
After Lockdown (After) &  9672/43263 (22.4\%) & Started after 01/06/2021. Ended before 01/04/2023 \\
\end{tabular}
\caption{HHSC population cohorts.}
\label{tb.cohorts}
\end{table}

% - - - - - - - - - - - - - - - - - - - - - - - - - - - - - - - - 
\subsection{Representing Shelter Access using Graphs}
\label{ssec:graphs}

A graph consists of a series of {\em nodes} connected by {\em edges} \parencite{bollobas1998}.  A {\em directed graph} associates a direction of travel from one node to the next.  Edges in a directed graph are typically represented using arrows.  A directed graph is created from the HHSC shelter data described in Section~\ref{ssec:data} where each emergency shelter is a node.  A directed edge is created between two nodes if one or more shelter users uses the first shelter and then makes a transition to use the second shelter.  

In addition to shelter nodes, the HHSC directed graph will also contain {\em gateway} nodes that represent entries to and exits from the HHSC.  The {\em Entry} gateway node marks the point where a person first enters the HHSC and the {\em Exit} node marks the point where a person makes a final exit from the HHSC and no longer appears in the data set.  The {\em Gap} gateway node represents when a person disappears from the data for a period of 30 days or longer but reappears at a later time to continue making use of the HHSC.  Interactions with a housing placement or other support program would also be recorded as a Gap if the person was not also actively using shelter at the same time.  A Multiple gateway node is also created to capture when a person interacts with multiple shelters within a 24 hour period.   Creating a composite node is primarily to simplify the graph visualization since people can access multiple shelters in a large number of unique combinations.  Representing each combination of shelter access by a unique node would cause an unacceptable amount of clutter in the graph visualization.

When visualizing a graph, it is useful to adjust the size of nodes and edges to convey information on relative utilization.  For example, a busier shelter can be represented using a bigger dot and a commonly used transition between shelters as a thicker arrow.  This utilization information can be determined simply by counting the total number of stays in a shelter and the total number of times a person makes a transition from one shelter to another.

Gaps in a person's record of shelter access are common and judging when a gap is large enough to represent a significant departure from an HHSC is subjective.  However, this study will adopt a threshold of 30 days which is commonly used to define different episodes of homelessness \parencite{kuhn1998, kneebone2015, aubry2013}.  This means that if there is a gap of 30 days or less between a person accessing first Shelter A and then Shelter B, that will be recorded as a direct transition from Shelter A to Shelter B.  If the gap is longer than 30 days, it would be recorded as a transition from Shelter A to the Gap node and then a second transition from the Gap node to Shelter B.

\begin{figure}[htbp]
\centerline{\includegraphics[width=3in]{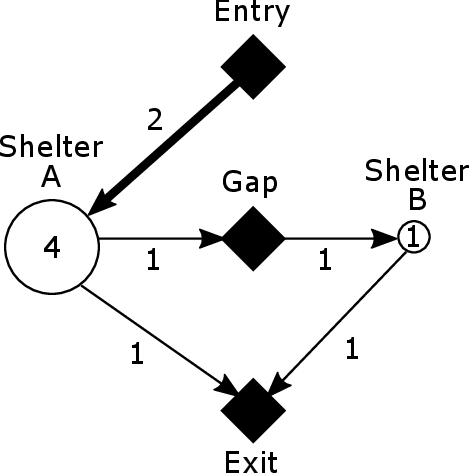}}
\caption{Example HHSC graph visualization.}
\label{fg.example}
\end{figure}

For example, consider an observation window where two people interact with an HHSC consisting of Shelter A and Shelter B.  The data shows that the first person used Shelter A for a single day.  This is recorded as one transition from Entry to Shelter A, one interaction with Shelter A and one transition from Shelter A to Exit.  The data shows the second person first using Shelter A for three consecutive days and then using Shelter B 31 days later for a single day.  This is recorded as a transition from Entry to Shelter A, three interactions with Shelter A, a transition from Shelter A to Gap to account for the 31 day absence, a transition from Gap to Shelter B, a single interaction with Shelter B and then a transition from Shelter B to Exit.  The directed graph showing HHSC utilization for these two people is shown in Figure~\ref{fg.example}.

% --------------------------------------------------------------
\section{Results}
\label{sec:results}

% - - - - - - - - - - - - - - - - - - - - - - - - - - - - - - - - 
\subsection{Shelter Stay and Transition Timelines}
\label{ssec:timelines}

\begin{figure}[htbp]
\centerline{\includegraphics[width=5in]{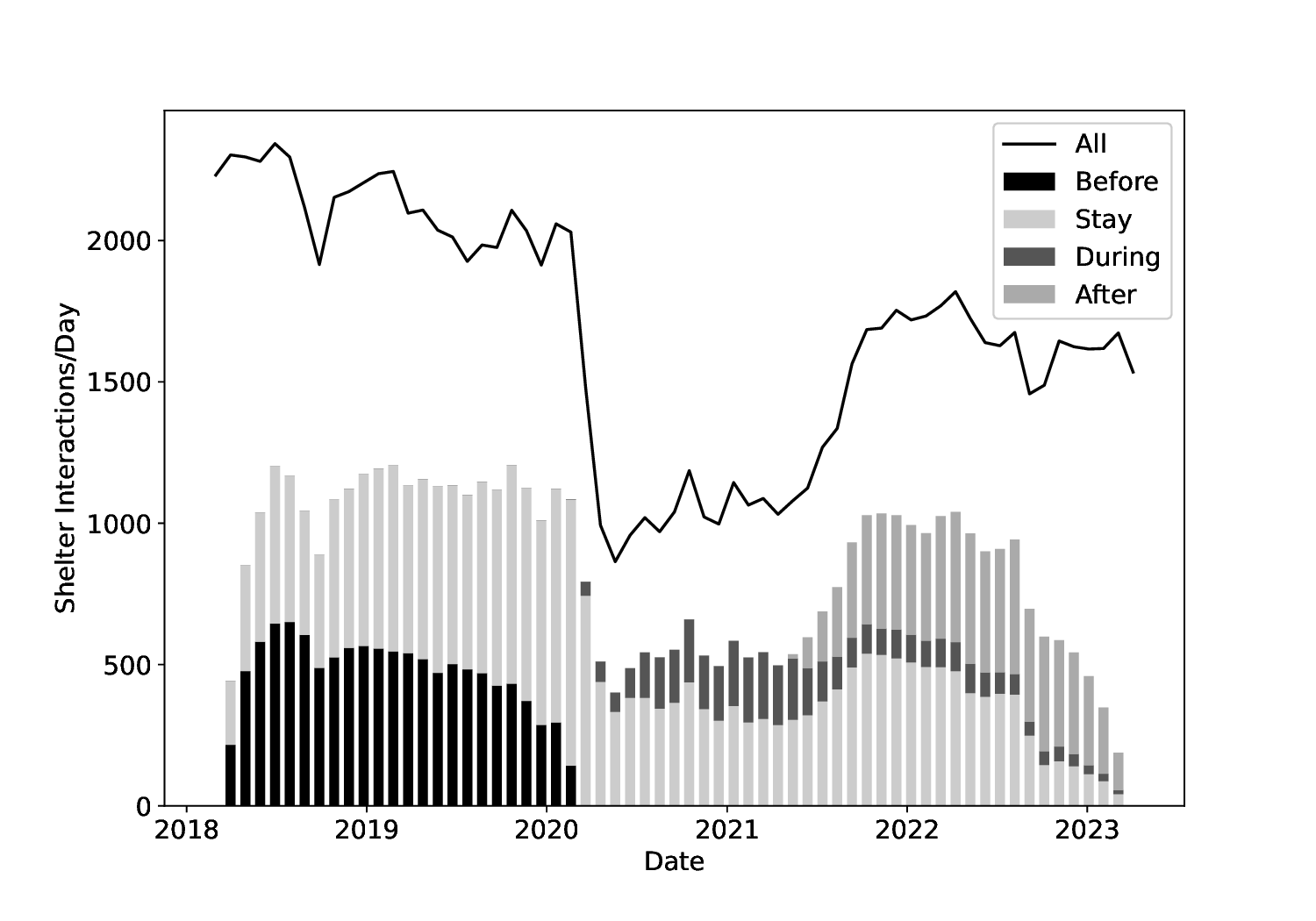}}
\caption{Shelter interaction timeline.}
\label{fg.accTL}
\end{figure}

\begin{figure}[htbp]
\centerline{\includegraphics[width=5in]{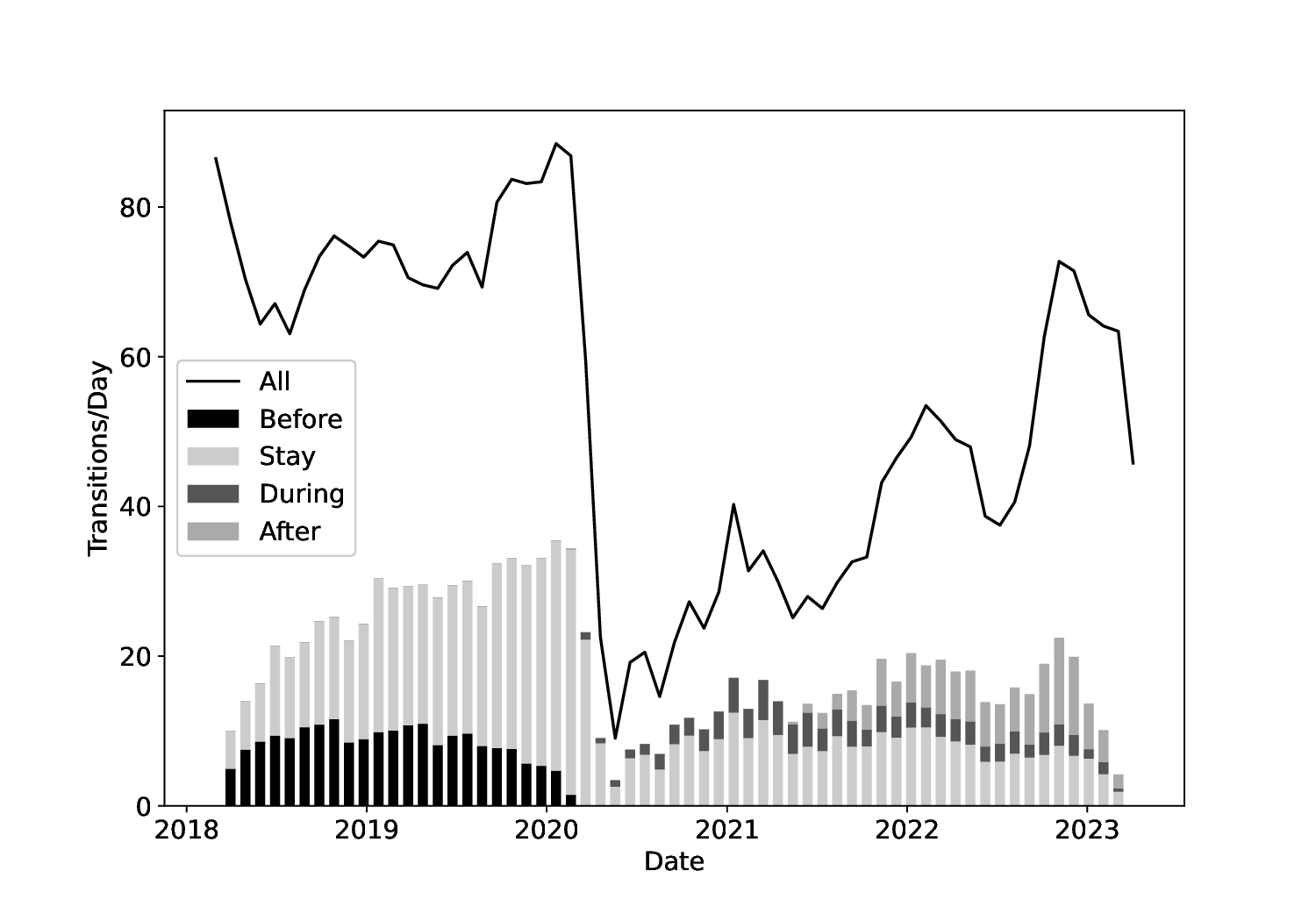}}
\caption{Shelter transition timeline.}
\label{fg.transTL}
\end{figure}

Figure~\ref{fg.accTL} shows the average number of times people interact with shelters in the Calgary HHSC from March 1, 2018 to May 1, 2023.  The figure also shows the number of interactions for each of the cohorts in Table~\ref{tb.cohorts}.  Note that average number of interactions per day will be higher than the more commonly reported number of unique individuals using the shelter system per day \parencite{calgaryhomelessfoundation2022} since Figure~\ref{fg.accTL} can include the same person accessing multiple shelters and/or day and night sleep services in the same day.  The figure also shows that shelter use by the Before cohort seems to taper in anticipation of lockdown.  However, this is an artifact of how the cohort is created.  The cohorts in Table~\ref{tb.cohorts} are selected to include entire records of shelter access.  Since only a minority of the Before cohort will have their records finish immediately before the start of lockdown, a tapering effect is observed.  Figure~\ref{fg.transTL} shows the average number of shelter transitions per day where a transition is calculated as described in Section~\ref{ssec:graphs}.

\begin{figure}[htbp]
\centerline{\includegraphics[width=5in]{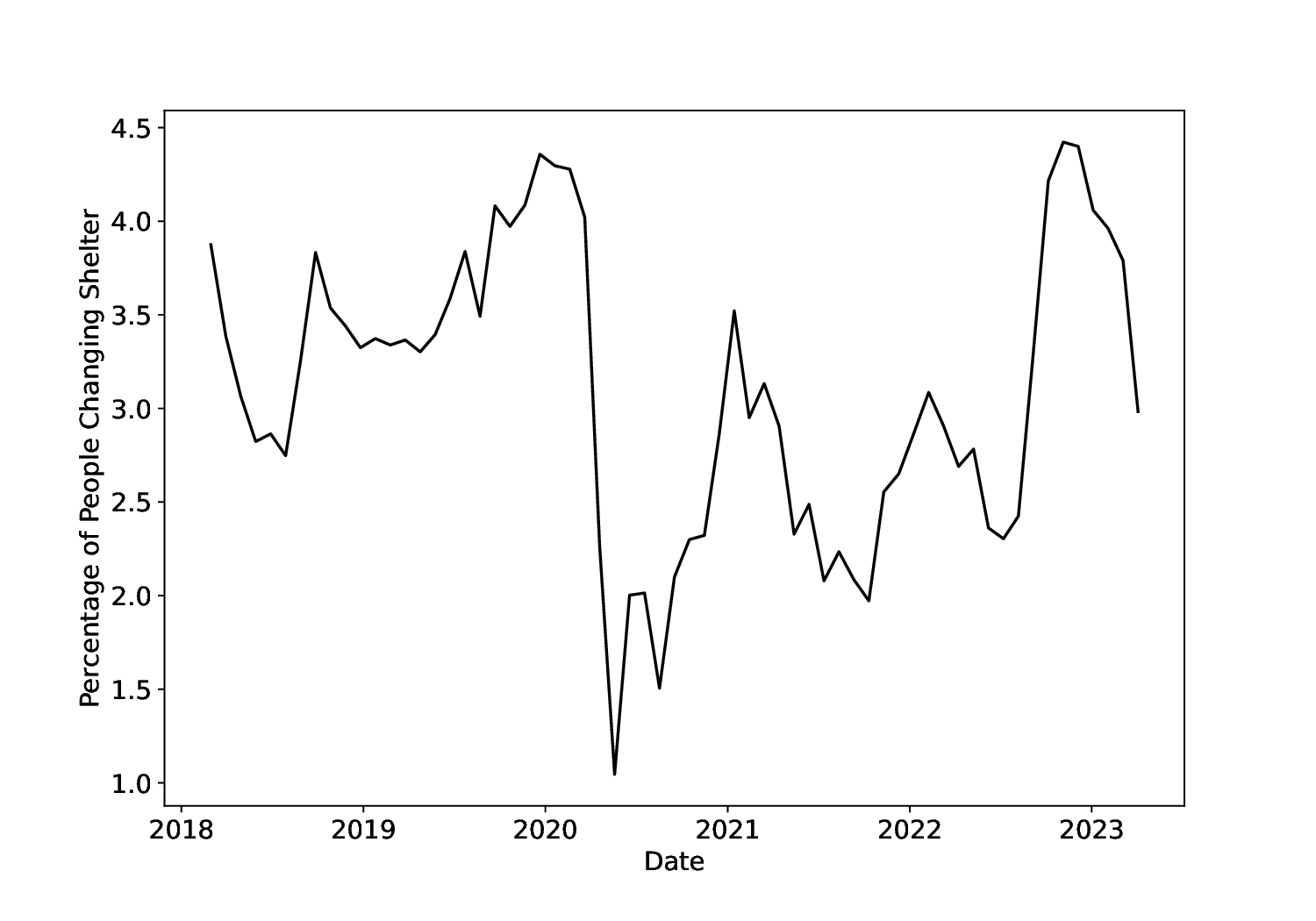}}
\caption{Proportion of shelter interactions that involve transition to a different shelter.}
\label{fg.transPct}
\end{figure}

Clearly, the total number of times shelter is accessed and the total number of shelter transitions reduce dramatically at the start of lockdown.  Both remain low until mid 2021.  However, since the purpose of this paper is to study how people move through a system of care, it is worth examining whether the reduction in transitions reflects a fundamental change in how shelters are used or if it is primarily an artifact of simply having fewer people in the system.  This can be established using Figure~\ref{fg.transPct} which shows the number of transitions per day divided by the number of shelter interactions per day.  This figure can be interpreted as the percentage of shelter interactions that involve a change to a different shelter.

The total number of transitions in Figure~\ref{fg.transTL} for the pre-lockdown era matches the results in \parencite{jadidzadeh2020} almost exactly.  While the shelter interaction values in Figure~\ref{fg.accTL} cannot be directly compared to the census style results in \parencite{jadidzadeh2020}, the authors of that paper do demonstrate that the overall percentage of people choosing to switch shelters on any particular day is quite low.  This is consistent with Figure~\ref{fg.transPct}.

% - - - - - - - - - - - - - - - - - - - - - - - - - - - - - - - - 
\subsection{Cohort Shelter Access Statistics}
\label{ssec:stats}

Clearly, the cohorts of people shown in Figures~\ref{fg.accTL} and \ref{fg.transTL} utilized the Calgary HHSC very differently before, during and after lockdown.  It is important to take a closer look at the individuals in each of these groups to better understand why.  Table~\ref{tb.stats} shows shelter access statistics for the cohorts in Table~\ref{tb.cohorts} during the pre-lockdown, lockdown and post-lockdown periods.  The table shows total number stays, the number of days between a person's first and last day using the HHSC (their {\em tenure}), their shelter use percentage (number of stays divided by tenure), number of unique shelters accessed and number of shelter to shelter transitions.  Shelter use percentage is included since it has been shown to be a good metric to differentiate between long term steady/chronic and long term sporadic/episodic shelter access patterns \parencite{messier2023}.

While some cohorts are present during multiple eras, the statistics shown for each cohort are calculated only using the portion of a person's record that falls within an era.  For example, pre-lockdown statistics for the Stay cohort during lockdown were calculated using only the portion of their HHSC access patterns that fell within the date range of the pre-lockdown period.

\begin{table}[htbp]
\centering
\begin{tabular}{r|cc|cc|ccc}
& \multicolumn{2}{|c|}{Pre-Lockdown} & \multicolumn{2}{|c|}{Lockdown} & \multicolumn{3}{|c}{Post-Lockdown} \\
& \multicolumn{2}{|c|}{Cohorts} & \multicolumn{2}{|c|}{Cohorts} & \multicolumn{3}{|c}{Cohorts} \\ \cline{2-8}
& Before & Stay & During & Stay & During & Stay & After \\ \hline
Period Duration (days) & 687 & 687 & 381 & 381 & 639 & 639 & 639 \\ \hline
Median Tenure (days) & 11 & 123 & 8 & 25 & 75 & 49 & 10 \\
Mean Tenure (days) & 87.6 & 209.7 & 37.6 & 79.4 & 161.5 & 124.2 & 61.8 \\
95th Pctl Tenure (days) & 448 & 612 & 192 & 335 & 552 & 473 & 317 \\ \hline
Median Stays & 4 & 19 & 4 & 9 & 11 & 13 & 4 \\
Mean Stays & 22.5 & 46.5 & 17.2 & 27.9 & 36.1 & 33.4 & 21.4 \\
95th Pctl Stays & 99 & 175 & 78 & 124 & 161 & 122 & 95 \\ \hline
Median Use Percent & 6.7\% & 15.6\% & 10.0\% & 16.7\% & 10.0\% & 16.8\% & 10.0\% \\
Mean Use Percent & 24.3\% & 27.2\% & 28.5\% & 30.6\% & 25.1\% & 30.5\% & 26.9\% \\
95th Pctl Use Percent & 100\% & 100\% & 100\% & 100\% & 100\% & 100\% & 100\% \\ \hline
Median Unique Shelters & 1 & 1 & 1 & 1 & 1 & 1 & 1 \\
Mean Unique Shelters & 1.1 & 1.2 & 1.1 & 1.2 & 1.3 & 1.2 & 1.1 \\
95th Pctl Unique Shelters & 2 & 2 & 2 & 2 & 3 & 2 & 2 \\ \hline
Median Transitions & 0 & 0 & 0 & 0 & 0 & 0 & 0 \\
Mean Transitions & 0.4 & 1.5 & 0.3 & 0.8 & 1.5 & 0.8 & 0.4 \\
95th Pctl Transitions & 2 & 8 & 2 & 4 & 7 & 4 & 2 \\ \hline
\end{tabular}
\caption{Shelter access statistics.}
\label{tb.stats}
\end{table}

% - - - - - - - - - - - - - - - - - - - - - - - - - - - - - - - - 
\subsection{Graph Visualization}
\label{ssec:viz}

\begin{figure}[htbp]
\centerline{\includegraphics[width=7in]{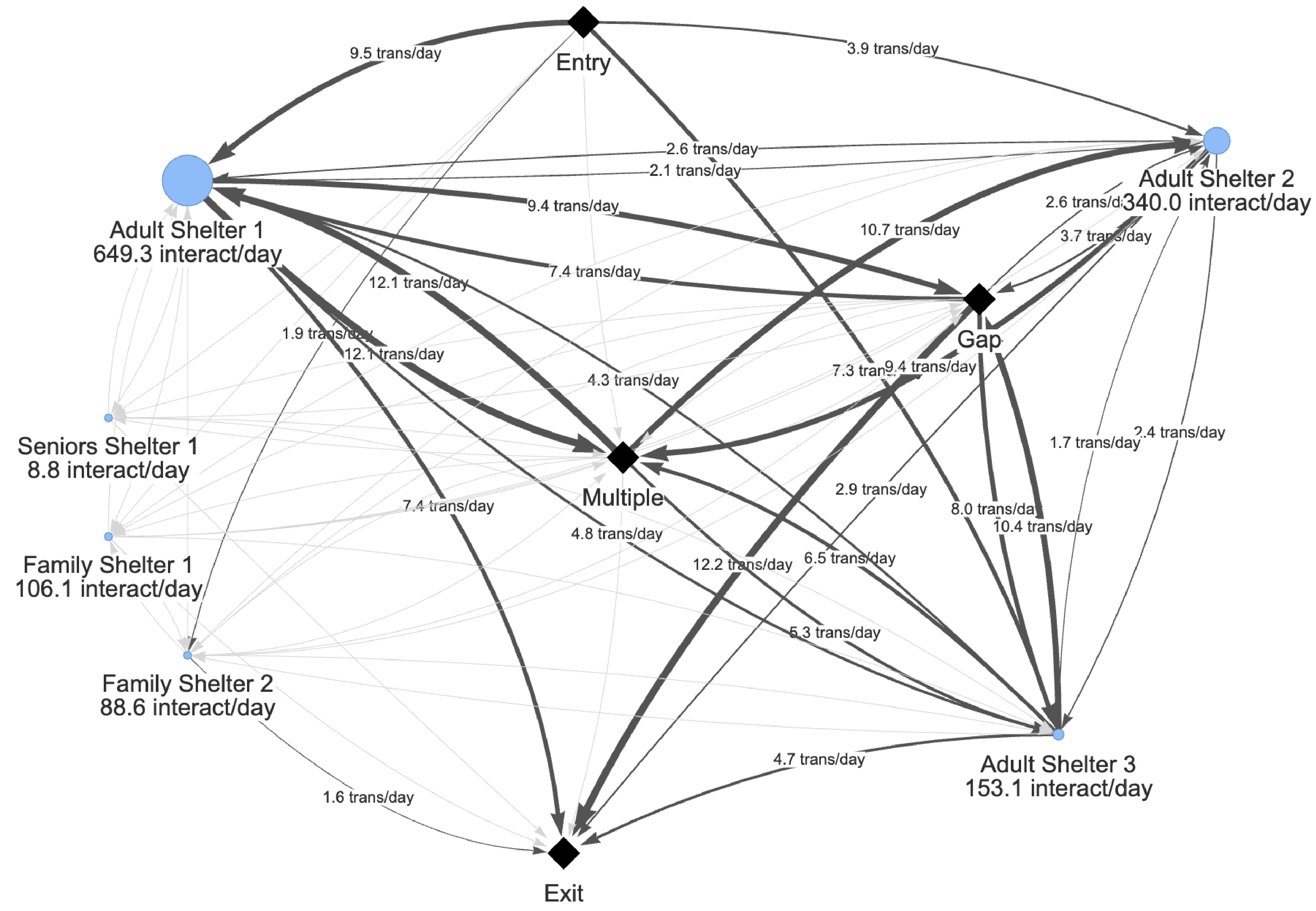}}
\caption{Calgary HHSC shelter utilization before lockdown.}
\label{fg.preLockdown}
\end{figure}

Figure~\ref{fg.preLockdown} shows a visualization of the directed graph representation of the Calgary HHSC described in Section~\ref{ssec:graphs} for all shelter users in the pre-lockdown period.  Rather than displaying an absolute count of shelter interactions and transitions, as shown in Figure~\ref{fg.example}, the interaction and transition counts were divided by the duration of the pre-lockdown era (687 days) to arrive at an average number of stays and transitions per day.  Edge thickness is proportional to the number of transitions per day except that edges representing fewer than one transition/day are unlabeled, shown in light grey and have a fixed width.

\begin{figure}[htbp]
\centerline{\includegraphics[width=7in]{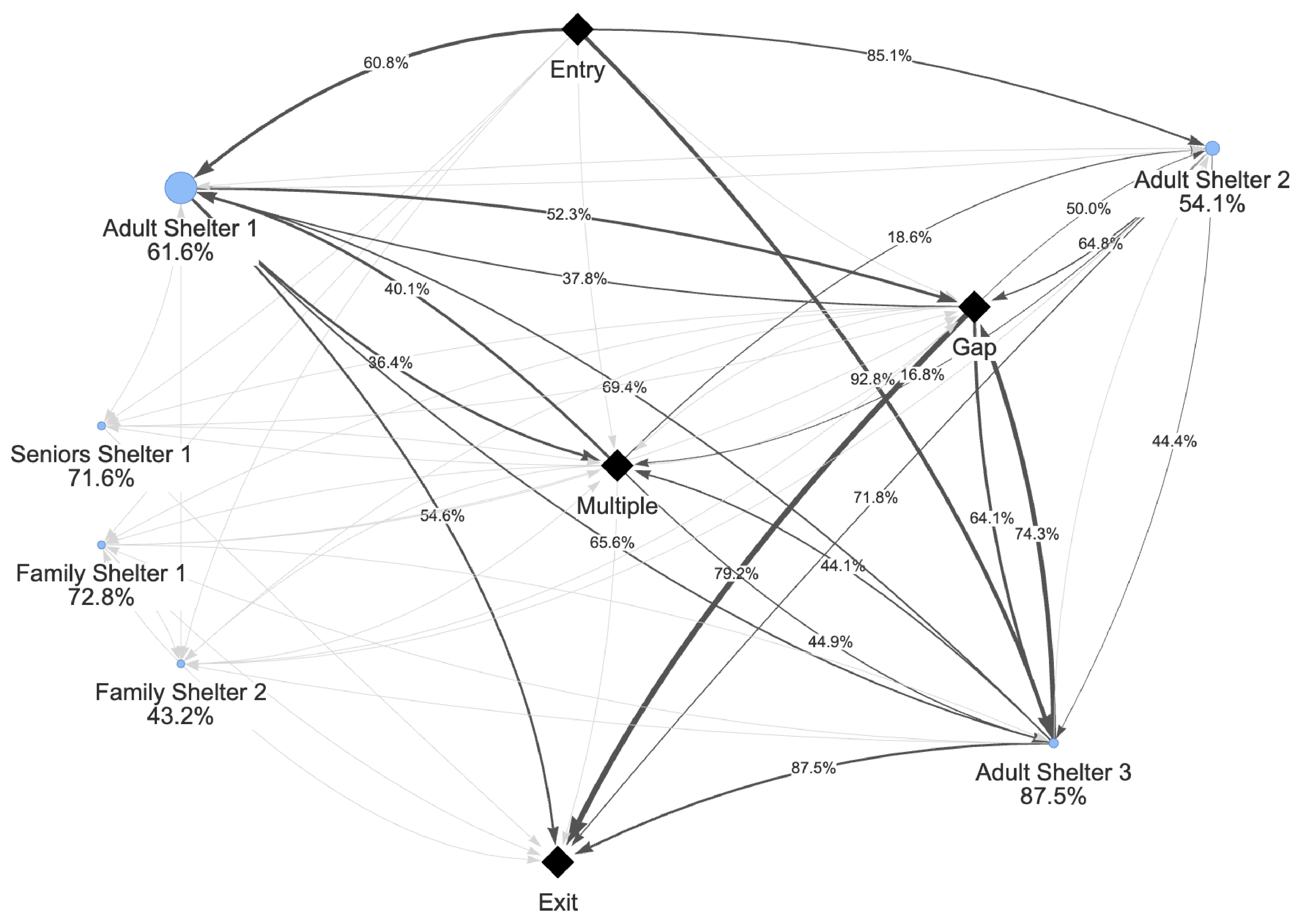}}
\caption{Calgary HHSC shelter utilization during lockdown.}
\label{fg.lockdown}
\end{figure}

\begin{figure}[htbp]
\centerline{\includegraphics[width=7in]{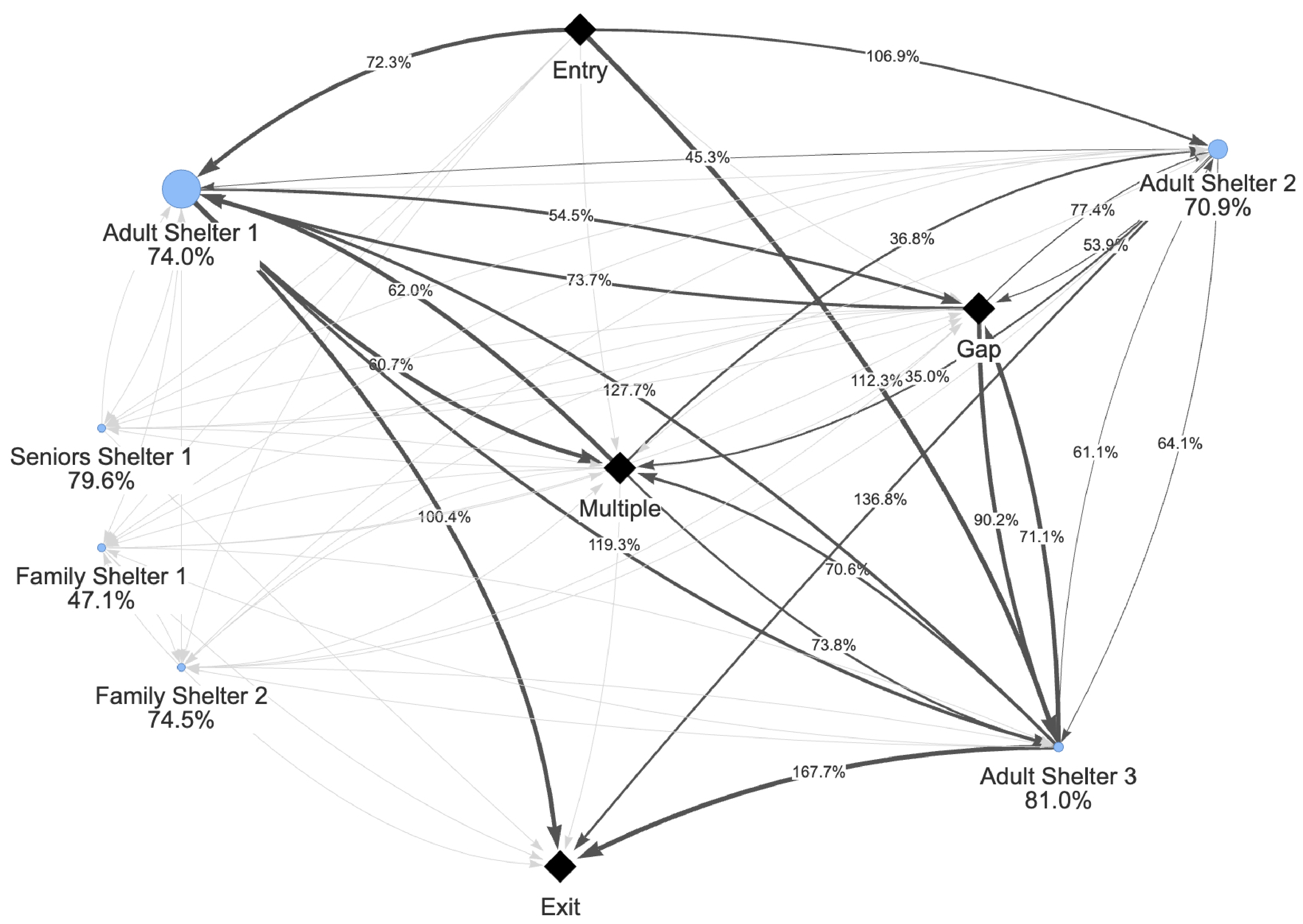}}
\caption{Calgary HHSC shelter utilization after lockdown.}
\label{fg.postLockdown}
\end{figure}

Figures~\ref{fg.lockdown} and \ref{fg.postLockdown} demonstrate shelter use and shelter transition frequencies during the lockdown and post-lockdown eras respectively.  Both shelter interaction and transition frequencies are expressed as percentages relative to the pre-lockdown era.  For example, in Figure~\ref{fg.lockdown}, the average number of people using the Adult Shelter 2 shelter per day was 54.1\% of the average number per day using that shelter during the pre-lockdown period.

% --------------------------------------------------------------
\section{Discussion}
\label{sec:discussion}

The timeline diagrams in Section~\ref{ssec:timelines} demonstrate that the drop in shelter use due to lockdown is remarkably well defined.  After approximately July 1, 2021, use of shelter climbs quickly to levels comparable to the pre-lockdown period, even though winter 2021 saw the arrival of the Omicron COVID-19 wave and periodic pandemic restrictions on gathering, workplaces and schools persisted until early 2022.  This return to shelter may have been encouraged by the uptake of the COVID-19 vaccine during the summer of 2021 and an increased comfort with using personal protective equipment (PPE) within shelter \parencite{calgaryherald2021}.

While Figures~\ref{fg.transTL} and \ref{fg.transPct} indicate that transitions between shelters did decrease during lockdown, Figure~\ref{fg.transPct} indicates that the transitions per shelter interaction remained relatively constant after an initial dip.  Comparing Figures~\ref{fg.accTL} and \ref{fg.transTL} also shows that the Stay cohort was responsible for a higher proportion of shelter transitions than the During cohort.  This suggests that longer term shelter users with established patterns of moving between shelter will continue those patterns even during emergency.  However, consistent with \parencite{jadidzadeh2020}, the most important message from Figure~\ref{fg.transPct} is that only a small percentage of shelter interactions overall involve a change to a different shelter.  This is important since high traffic between shelters during a pandemic would be a mechanism to quickly spread infection from one shelter to the next.

Figure~\ref{fg.accTL} also shows that the Stay cohort represents a group of people who were making heavy use of shelter before lockdown and continued to rely heavily on the shelter system after the pandemic began.  This is supported by Table~\ref{tb.stats} which shows the Stay cohort using shelter much more frequently than the cohort who exited the system before lockdown or the cohort who used shelter for the first time during lockdown.  It is intuitive that a group of people making heavy use of shelter will have few options other than to continue relying on shelter even during an emergency that would make congregate living much more hazardous.  However, Figure~\ref{fg.accTL} and Table~\ref{tb.stats} are a reminder than any planning for a future pandemic response should anticipate continued shelter use by those who are currently interacting with shelter on a consistent basis.

Table~\ref{tb.stats} demonstrates that it is very uncommon for a person to interact with a large number of shelters.  Even the 95th percentile of all the cohorts and periods shown in Table~\ref{tb.stats} interact with 2 or 3 unique shelters.  The median number of transitions is 0 and only the cohorts with well established records of shelter interactions have a high number of transitions in the 95th percentile.  However, the number of unique shelters statistics suggest that even these higher number of transitions are most likely switching back and forth between the same two shelters.  

In Table~\ref{tb.stats}, it is important to note that the cohort in the post-lockdown era most likely to make shelter transitions are those who started using shelter for the first time during lockdown.  Accessing shelter for the first time during a pandemic would be a daunting decision.  A person making this choice would very likely be facing multiple challenges that would have left them with few other options.  Now in the post-lockdown period, Table~\ref{tb.stats} suggests that this group displays a longer term, sporadic shelter access pattern with a relatively higher number of shelter transitions.  This is unusual and may identify a unique cohort of people who warrant special attention both by researchers and support agencies within the HHSC.

While Sections~\ref{ssec:timelines} and \ref{ssec:stats} provide information on aggregate shelter interaction and what people experience on the individual level, the visualizations presented in Section~\ref{ssec:viz} demonstrate the highly complex and interconnected nature of the Calgary HHSC when we visualize the unique sources and destinations associated with shelter transitions. Figure~\ref{fg.preLockdown} shows that the traffic into and out of each shelter is proportional to shelter size (ie. bigger shelters see more traffic).  There is also considerable traffic with the Gap node due to the episodic nature of how many people interact with emergency shelters.  Interestingly, the traffic into/out of the Multiple shelter node (indicating interactions with multiple shelters in one day) is approximately comparable to the sum of the smaller arrows representing direct transition between shelters.  This means that a sizable proportion of inter-shelter interactions occur in very close proximity.

The network visualization in Figure~\ref{fg.preLockdown} also provides insight into the nature of the people accessing particular shelters.  For example, the traffic between the Gap node and Adult Shelter 3 is comparable the traffic between Gap and the Adult Shelter 1 even though the Adult Shelter 1 is considerably larger.  This suggests that the people using Adult Shelter 3 are more episodic in how they access shelter which is important information when designing programs for Adult Shelter 3 users.

Figure~\ref{fg.lockdown} shows the overall expected decrease in shelter interactions and transitions for the lockdown period.  In many cases, the a shelter’s transition traffic and the number of interactions with that shelter decrease by approximately the same amount. For example, Adult Shelter 1 shelter interactions drop to 61.6\% pre-lockdown levels and the traffic from Entry to Adult Shelter 1 and Adult Shelter 1 to exit decrease to 60.8\% and 54.6\%, respectively. However, there are also some interesting variations. Many transitions directly between shelters and to/from the Multiple node decrease by larger margins than the shelter interactions. This suggests a tendency to “shelter-in-place” during lockdown.

Figure~\ref{fg.postLockdown} shows post-lockdown shelter use approaching but still below pre-lockdown utilization.  This is mainly because the post-lockdown era, as defined in this paper, does include approximately one year of shelter operations where COVID-19 restrictions would have had at least some effect on shelter and program capacity.  The picture is also not uniform.  For example, Adult Shelter 3 traffic is at or slightly above pre-lockdown levels.  Figure~\ref{fg.postLockdown} also shows a higher rate of direct transition to system exit from the top three busiest shelters: 100.4\%, 136.8\% and 167.7\% for the Adult Shelter 1, Adult Shelter 2 and Adult Shelter 3, respectively.  While there are many reasons for leaving an HHSC, this increase in system exits does co-incide with shelters continuing to make transitioning people from shelter to housing a priority in the post-pandemic era.

% --------------------------------------------------------------
\section{Conclusion}
\label{sec:conclusion}

The pandemic had a profound impact on the operation of HHSCs and the people they serve.  Understanding this impact in retrospect provides valuable information on how an HHSC can be better prepared to support people during future pandemics or other emergencies.

The visualizations presented in Section~\ref{ssec:viz} reveal an HHSC consisting of agencies that are richly connected by the journeys of the people they serve.  Almost every possible transition between agencies was observed.  Figure~\ref{fg.transPct} also indicates that the proportion of people choosing to move between shelters remains surprisingly robust over the pandemic timeline.  At the same time, this same figure shows that the overall percentage of shelter interactions that involve a transition to a different shelter remain quite small.  This means that monitoring or attempting to discourage transitions between shelter in future pandemics may not be the best use of resources.

A higher impact activity would be to better understand the people who continue to rely on emergency shelters during future pandemics or other emergencies.  Not surprisingly, Table~\ref{tb.stats} shows that these people tend to be the longer term or chronic shelter users.  These people require different supports than short term or episodic shelter users and their voices should be included in any future HHSC pandemic response plans.

Finally, characterizing an HHSC as a directed graph is a powerful tool for understanding the complex and highly connected way people access the agencies making up the HHSC.  Pandemic or not, system flow visualizations like those presented in Section~\ref{ssec:viz} reveal how agencies are connected by the people they serve.  This perspective should be a powerful motivator for the operators of those agencies to continue to collaborate and communicate along the lines of how people flow into and out of their programs.  It also encourages us to look at an HHSC as single system of care since that is how many people choose to engage with it.

% --------------------------------------------------------------
\section{Acknowledgments}
\label{sec:ack}

The author would like to gratefully acknowledge the support of Making the Shift, the Calgary Homeless Foundation and the Government of Alberta.  This study is based in part on data provided by Alberta Seniors, Community and Social Services.  The interpretation and conclusions contained herein are those of the researchers and do not necessarily represent the views of the Government of Alberta.  Neither the Government of Alberta nor Alberta Seniors, Community and Social Services express any opinion related to this study.

\printbibliography

\end{document}